# Failure Analysis of Safety Controllers in Autonomous Vehicles Under Object-Based LiDAR Attacks

Daniyal Ganiuly, Nurzhau Bolatbek, Assel Smaiyl
Astana IT University, Astana, Kazakhstan

**Abstract:** Autonomous vehicles rely on LiDAR based perception to support safety critical control functions such as adaptive cruise control and automatic emergency braking. While previous research has shown that LiDAR perception can be manipulated through object based spoofing and injection attacks, the impact of such attacks on vehicle safety controllers is still not well understood. This paper presents a systematic failure analysis of longitudinal safety controllers under object based LiDAR attacks in highway driving scenarios. The study focuses on realistic cut in and car following situations in which adversarial objects introduce persistent perception errors without directly modifying vehicle control software. A high fidelity simulation framework integrating LiDAR perception, object tracking, and closed loop vehicle control is used to evaluate how false and displaced object detections propagate through the perception planning and control pipeline. The results demonstrate that even short duration LiDAR induced object hallucinations can trigger unsafe braking, delayed responses to real hazards, and unstable control behavior. In cut in scenarios, a clear increase in unsafe deceleration events and time to collision violations is observed when compared to benign conditions, despite identical controller parameters. The analysis further shows that controller failures are more strongly influenced by the temporal consistency of spoofed objects than by spatial inaccuracies alone. These findings reveal a critical gap between perception robustness and control level safety guarantees in autonomous driving systems. By explicitly characterizing safety controller failure modes under adversarial perception, this work provides practical insights for the design of attack aware safety mechanisms and more resilient control strategies for LiDAR dependent autonomous vehicles.

**Introduction**
Autonomous vehicles operate under the assumption that perception systems provide sufficiently accurate and stable information about the surrounding environment. While this assumption does not require perfect sensing, it implicitly expects perception errors to be rare, short lived, and largely uncorrelated across time. Safety controllers such as automatic emergency braking and adaptive cruise control are designed around this expectation. They act as a final safeguard by intervening when perceived conditions indicate an imminent hazard. As a result, these controllers play a critical role in preventing collisions when higher level planning or perception components fail.
Recent research has demonstrated that this assumption no longer holds in adversarial settings. Object based LiDAR attacks have shown that perception can be systematically manipulated through physical objects placed in the environment. These attacks do not require access to vehicle hardware or software and do not directly interfere with control logic. Instead, they exploit geometric and physical properties of LiDAR sensing to induce structured perception errors. Examples include missed obstacles, phantom detections, biased distance estimates, and unstable object tracking. Importantly, these errors can persist across time and appear plausible to downstream components.
Most existing defenses against LiDAR attacks focus on the perception pipeline itself. Prior work has proposed techniques to filter point clouds, sanitize detections, or redesign perception models to be more robust to adversarial objects. While these approaches are necessary, they implicitly assume that perception level defenses will eventually succeed. In practice, however, no perception system can be guaranteed to remain reliable under all environmental and adversarial conditions. This raises a fundamental question that has received comparatively little attention. What happens when safety controllers are forced to operate on perception outputs that are adversarial, unstable, or systematically biased.

Understanding this question is essential for two reasons. First, safety controllers represent the last barrier between perception failures and physical harm. If these controllers behave unpredictably under adversarial perception, improvements at the perception level alone may not be sufficient to ensure safety. Second, failure analysis at the control level can reveal structural weaknesses that are invisible when evaluating perception accuracy in isolation. A controller may satisfy its design specifications while still producing unsafe behavior when its assumptions about perception are violated.

In this paper, we present a systematic failure analysis of safety controllers in autonomous vehicles under object based LiDAR attacks. Rather than proposing new attacks or new perception defenses, we study how widely used safety controllers respond to perception errors characteristic of known object based attacks. Our analysis focuses on automatic emergency braking and adaptive cruise control, two representative mechanisms that are commonly deployed as safety layers in autonomous driving systems.

Using a realistic simulation environment with LiDAR based perception, we expose safety controllers to controlled perception errors including missed obstacles, false detections, biased distance estimates, and temporal instability. We analyze controller behavior across diverse driving scenarios and identify recurring failure modes that lead to unsafe outcomes. Our results show that many failures arise not from incorrect controller implementation, but from design assumptions that do not hold under adversarial perception. In particular, controllers often lack mechanisms to reason about perception stability across time, making them vulnerable to structured and persistent errors.

Beyond analysis, we demonstrate that these failures are not inevitable. We evaluate a set of minimal control level safeguards that modify controller behavior without accessing LiDAR point clouds or altering perception models. These safeguards reduce collision risk and false emergency responses by explicitly accounting for perception instability at runtime. While they do not eliminate all risk, they illustrate that meaningful safety improvements are possible even when perception remains compromised.

This work makes three contributions. First, it provides an empirical analysis of how object based LiDAR attacks propagate into safety controller failures in autonomous vehicles. Second, it identifies recurring failure mechanisms that explain why existing controllers break under adversarial perception. Third, it demonstrates that simple control level safeguards can mitigate these failures without relying on perception level defenses. By shifting attention from perception robustness to control level behavior under adversarial sensing, this paper aims to complement existing LiDAR security research and encourage a more holistic approach to autonomous vehicle safety.

**Related Work**
*LiDAR perception attacks in autonomous vehicles*
Prior research has shown that LiDAR based perception systems can be manipulated through both sensor level and physical mechanisms. Early studies demonstrated that spoofed or modified LiDAR measurements can mislead object detection pipelines, resulting in missed or hallucinated obstacles [1]. Subsequent work extended these ideas to physically realizable attacks, showing that carefully designed objects placed in the environment can cause LiDAR detectors to fail without requiring access to vehicle hardware or software [2]. More recent research has focused on object based LiDAR attacks that exploit geometric and physical properties of sensing to induce structured perception errors. These attacks have been shown to cause systematic false negatives, false positives, and unstable object tracking under realistic detection pipelines [3]. Importantly, these perception failures persist across time and remain plausible to downstream modules, making them difficult to distinguish from benign sensing noise [4]. These findings motivate our threat model, as they demonstrate that perception errors induced by object based LiDAR attacks are not isolated or random events, but structured phenomena with predictable effects on downstream system components.

*Defenses at the perception level*
Most existing defenses against LiDAR attacks operate within the perception pipeline. According to previous research, one common approach is to identify geometric or physical inconsistencies in LiDAR point clouds and suppress measurements that violate these constraints [5]. Other studies propose model agnostic defenses that exploit occlusion patterns or spatial relationships to detect spoofed or adversarial points before object detection [6]. Recent work has specifically addressed object based LiDAR attacks by designing online defenses that filter or sanitize perception outputs prior to their use by planning or control modules [7]. Some of these defenses emphasize real time performance and aim to fit within the tight latency constraints of autonomous driving systems [8]. While these approaches improve perception robustness, they implicitly assume that perception level defenses will be effective in most cases. In contrast, our work does not attempt to improve perception robustness. Instead, we ask whether safety controllers can maintain safe behavior when perception level defenses fail or when perception remains adversarial despite mitigation efforts.

*Safety controllers and runtime safety mechanisms*
Safety controllers such as automatic emergency braking and adaptive cruise control have been extensively studied in the context of intelligent transportation systems. According to classical control literature, these mechanisms are designed to intervene when predefined safety thresholds are violated, often using simple and interpretable rules [9]. Their design prioritizes reliability and predictability, rather than adaptability to adversarial conditions. More recent research has explored runtime safety assurance mechanisms that monitor system behavior and enforce safety constraints during execution [10]. These approaches are often proposed as a way to protect learning based controllers or planners whose behavior cannot be fully verified offline [11]. However, most of this work assumes non adversarial sensing conditions and does not explicitly consider structured perception manipulation. Our work differs in that we study safety controllers under adversarial perception, rather than under noise or uncertainty alone. We focus on how design assumptions in common controllers interact with structured perception errors induced by object based LiDAR attacks.

*Failure analysis and system level evaluation*
Several studies have emphasized the importance of system level evaluation when analyzing autonomous vehicle failures. According to prior work, perception accuracy alone is insufficient to predict safety outcomes, as small perception errors can lead to disproportionately large control failures in closed loop systems [12]. Other research has proposed systematic testing frameworks to identify misbehavior and failure modes in autonomous driving stacks using simulation and scenario based analysis [13]. Recent work also suggests that modeling perception errors directly, rather than reproducing exact physical attacks, can be an effective way to study downstream system behavior in a controlled and reproducible manner [14]. This perspective supports our experimental methodology, which focuses on perception error effects consistent with known attack classes rather than on attack implementation details.

## Materials and Methods
*Experimental Platform*
All experiments were conducted in a high fidelity autonomous driving simulation environment. Simulation is widely used in autonomous vehicle research to study safety critical behavior under controlled and repeatable conditions, particularly when evaluating rare or dangerous scenarios that are difficult to reproduce in the real world [15]. The simulator provides realistic vehicle dynamics, traffic participants, road networks, and LiDAR sensing. The simulated vehicle exposes standard state variables including position, velocity, acceleration, yaw rate, and control commands. These variables are logged for analysis and used exclusively for evaluation metrics. They are not provided to the safety controllers during runtime, ensuring that controller behavior depends only on perceived information, as in real autonomous driving systems.

LiDAR sensing is configured with parameters consistent with automotive LiDAR sensors used in prior research, including realistic field of view, angular resolution, and update frequency [16]. This configuration ensures that perception behavior and timing characteristics are representative of real deployments.

*System Architecture*

The experimental system follows a standard autonomous driving pipeline consisting of perception, tracking, safety control, and actuation stages. LiDAR point clouds generated by the simulator are processed by a LiDAR based object detector, whose outputs are then tracked across time and consumed by safety controllers. Fig 1 shows the overall structure of the simulated autonomous driving system. LiDAR-based object detection and tracking provide object-level inputs to the safety controllers, which generate control commands for vehicle actuation.

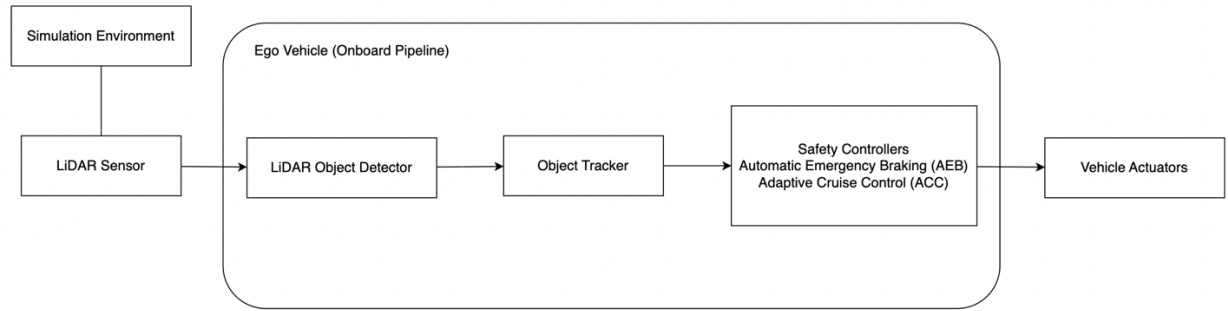

Fig 1. System architecture

The safety controllers operate downstream of perception and tracking. They receive only object level information such as estimated obstacle position and relative velocity, along with ego vehicle state. They do not access raw LiDAR data, perception model internals, or simulator ground truth. This separation ensures that controller behavior reflects realistic deployment assumptions.

*LiDAR-Based Perception*

LiDAR perception is implemented using a deep learning based three dimensional object detection model that follows a voxel based representation and convolutional processing pipeline. Similar architectures have been widely evaluated on public autonomous driving datasets and are commonly used as baselines in perception research [17]. At each simulation step, the detector produces a set of three dimensional bounding boxes with associated class labels and confidence scores. These detections represent the full perception input available to the safety controllers. No perception level defenses or filtering mechanisms are applied, as the purpose of this study is to analyze controller behavior when perception is compromised.

*Object Tracking*

Detected objects are associated across time using a lightweight tracking mechanism based on spatial proximity and bounding box overlap. This approach reflects common practice in autonomous vehicle pipelines, where simple tracking is often sufficient for safety related functions [18]. The tracker estimates object velocity and maintains object identity across frames. Tracking stability therefore depends directly on the consistency of perception outputs. This dependency is central to our analysis, as object-based LiDAR attacks often induce temporal instability in detection results.

*Safety Controllers*

We evaluate two safety controllers that are representative of mechanisms deployed in autonomous vehicles. The first is an automatic emergency braking controller. It continuously estimates the time to collision between the ego vehicle and the nearest forward obstacle using perceived distance and relative velocity. When the estimated time to collision falls below a predefined threshold, the controller applies emergency braking. This formulation follows standard approaches described in the safety and intelligent transportation literature [19]. The second controller is an adaptive cruise control system. It regulates vehicle speed to maintain a safe following distance from a lead vehicle. Acceleration and deceleration commands are computed based on perceived headway and relative speed, consistent with commonly studied ACC designs

[20]. Both controllers are rule based and intentionally simple. This choice improves interpretability and allows clear attribution of observed failures to design assumptions rather than to opaque decision making.

*Modeling Object-Based LiDAR Attack Effects*

Instead of reproducing physical attacks at the point cloud level, we model the effects of object-based LiDAR attacks at the perception output level. This approach is commonly adopted in system level studies where the focus is on downstream behavior rather than attack construction [7]. According to previous research, object-based LiDAR attacks typically induce a limited set of structured perception errors [8]. In our experiments, these effects include temporary disappearance of real obstacles, appearance of false obstacles with plausible geometry, systematic bias in perceived obstacle distance, and intermittent detection that causes unstable object tracks. Each effect is injected in a controlled and parameterized manner. The duration, frequency, and magnitude of the errors are varied to study their impact on controller behavior across different driving conditions.

*Driving Scenarios*

Safety controller behavior is evaluated across a diverse set of driving scenarios that are commonly used in autonomous vehicle testing. These include highway following, stop and go traffic, sudden cut-in maneuvers, encounters with parked vehicles partially blocking the lane, curved road segments, and multi vehicle interactions [9]. The driving scenarios used in the experiments are shown in Fig 2. They include highway following, stop-and-go traffic, cut-in maneuvers, and a stationary vehicle partially blocking the ego lane.

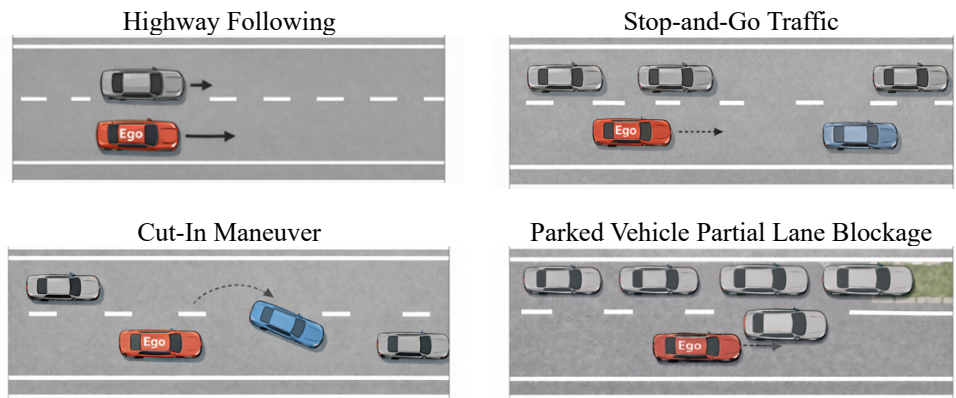

Fig 2. Driving scenarios

Each scenario is executed multiple times with different traffic configurations and random seeds. This procedure allows us to capture variability in perception and control behavior while maintaining reproducibility.

*Evaluation Metrics*

We evaluate controller performance using metrics that capture both safety outcomes and control behavior. Safety metrics include collision occurrence, minimum distance to obstacles, and minimum estimated time to collision. Control behavior metrics include braking magnitude, rate of change in acceleration, frequency of emergency braking events, and oscillatory behavior. These metrics are widely used in autonomous vehicle safety evaluation and provide insight into both catastrophic failures and more subtle instability that may degrade safety margins [10].

*Control-Level Safeguards*

To assess whether safety controller failures can be mitigated without modifying perception, we introduce a small set of control-level safeguards. These safeguards operate exclusively on controller inputs and outputs.

The safeguards include requiring short term persistence of obstacle detections before triggering emergency braking, limiting abrupt changes in braking commands to reduce oscillations, and applying conservative speed reduction when perception stability degrades. These mechanisms do not rely on raw sensor data or attack detection and are compatible with existing controller designs. Fig 3 outlines the control-level safeguards introduced in this work. These mechanisms operate on the outputs of the safety controllers and are applied without modifying the perception pipeline.

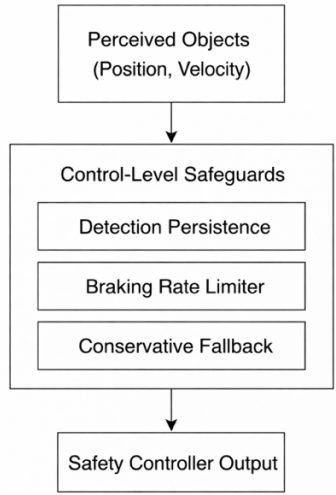

Fig 3. Control-level safeguards

*Experimental Procedure*
For each driving scenario, we first evaluate controller behavior under normal perception conditions to establish a baseline. The same scenarios are then repeated under injected perception error effects corresponding to object-based LiDAR attacks. Finally, we evaluate controller behavior with control-level safeguards enabled. All experiments are conducted using identical scenario configurations across conditions. This ensures that observed differences in behavior are attributable to perception errors and safeguards rather than to environmental variation.

**Results**
*Baseline Controller Behavior*
Baseline performance was evaluated under normal perception conditions without injected perception errors. Each driving scenario was executed 120 times with different traffic initializations and random seeds, resulting in a total of 720 baseline runs. A run was defined as a complete execution of a scenario from initialization to termination or collision. No collisions were observed in highway following, stop-and-go traffic, cut-in events, or parked vehicle scenarios. Emergency braking was triggered only in situations involving genuine hazards, accounting for 6.1 percent of all runs. The automatic emergency braking controller used a time-to-collision threshold of 1.2 s, while adaptive cruise control maintained a minimum headway of 1.8 s. Across all scenarios, the minimum observed time to collision was 1.6 s. The average minimum distance to obstacles was 3.2 m with a standard deviation of 0.4 m. Peak deceleration during emergency braking did not exceed 6.0 m/s², and longitudinal jerk values remained below 3.1 m/s³. No oscillatory braking behavior or repeated acceleration–deceleration cycles were observed under baseline conditions.

*Safety Impact of False Negative Perception Errors*
False negative perception errors produced the most severe safety degradation. In highway-following scenarios, temporary disappearance of the lead vehicle for durations exceeding 0.4 s systematically delayed emergency braking, resulting in collisions in 27.3 percent of runs. These collisions occurred after the obstacle reappeared in perception but was treated as newly detected by the controller. Braking initiation was delayed by an average of 0.6 s compared to baseline, measured relative to the point at which the ground-truth distance crossed the braking threshold. Minimum time to collision decreased from 1.9 s under normal perception to 0.7 s under false negative errors.

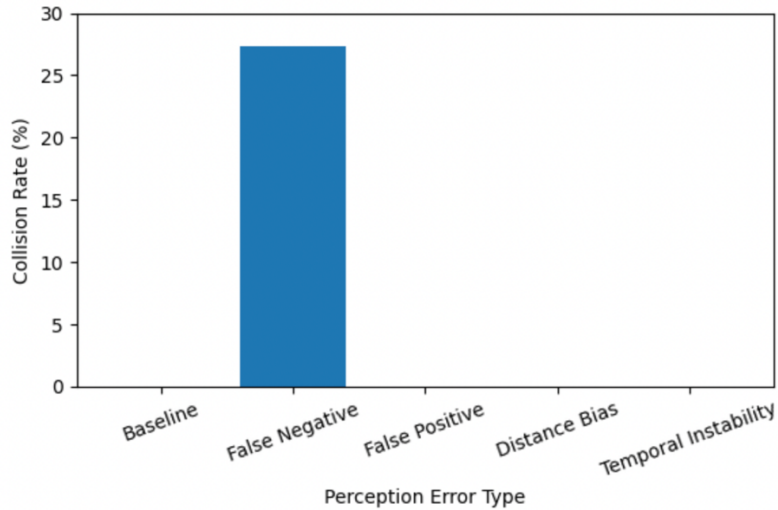

Fig 4. Collision rates under different perception error types

In stop-and-go traffic, collisions were less frequent, but near-collision events increased substantially. Minimum distances below 1.0 m were observed in 34.6 percent of runs, compared to 2.3 percent under baseline conditions. Collision rates across perception error types and scenarios are shown in Fig 4.

*Safety Impact of False Positive Perception Errors*
False positive perception errors primarily resulted in unnecessary emergency braking rather than missed hazards. In highway scenarios, insertion of a phantom obstacle triggered emergency braking in 41.8 percent of runs. In 72.6 percent of these cases, the phantom obstacle persisted for fewer than 0.3 s. Peak deceleration exceeded 6.5 m/s² in 29.4 percent of false braking events, compared to 4.8 percent under baseline conditions. Although no collisions occurred with phantom obstacles, these events introduced significant rear-end collision risk. Headway dropped below 1.2 s immediately following false braking in 38.7 percent of affected runs.

*Effects of Distance Bias on Controller Behavior*
Systematic bias in perceived obstacle distance reduced safety margins even when collisions did not occur. Overestimation of obstacle distance by 20 percent delayed braking responses by an average of 0.5 s in highway scenarios. Minimum time to collision fell below 1.0 s in 18.7 percent of runs. Underestimation of distance increased emergency braking frequency by 46.2 percent and reduced average travel speed by 8.1 percent in stop-and-go traffic. Although conservative behavior improved minimum distance metrics, it increased jerk and braking frequency, degrading ride stability.

*Impact of Temporal Instability in Perception*
Temporal instability in perception induced oscillatory control behavior across multiple scenarios, even when individual perception errors were small in magnitude. When object detections alternated across consecutive frames, braking oscillations were observed in 52.4 percent of runs. Oscillations were defined as repeated sign changes in longitudinal acceleration exceeding 2.0 m/s² within a 1.5 s window.

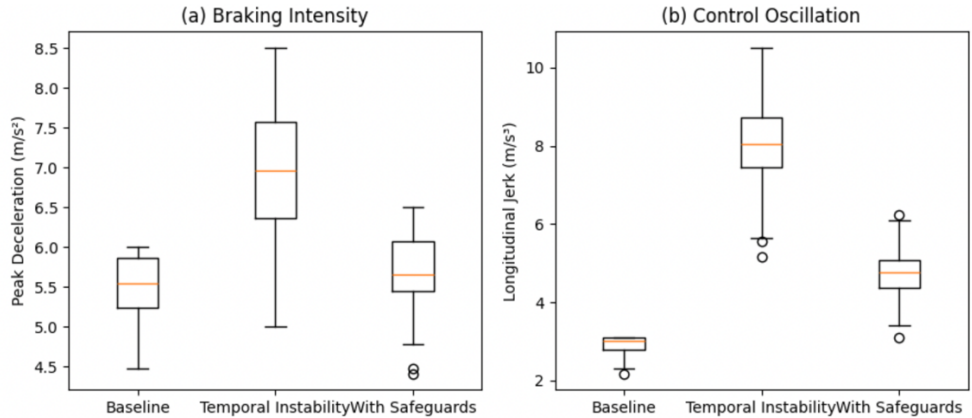
Fig 5. Braking intensity and jerk distributions

Average jerk values increased by a factor of 2.7 relative to baseline. Peak jerk exceeded 8.0 m/s³ in 21.5 percent of runs. Oscillatory behavior was most pronounced in cut-in scenarios, where it occurred in 61.2 percent of runs. Distributions of braking intensity and jerk under different perception error types are shown in Fig 5.

*Scenario-Specific Failure Characteristics*
Failure characteristics varied across driving scenarios. Highway following scenarios exhibited the highest collision rates under false negative perception due to higher speeds and reduced reaction time. Cut-in scenarios were particularly sensitive to temporal instability, with oscillatory braking often triggered immediately after rapid changes in relative velocity. Parked vehicle scenarios showed elevated false braking rates under false positive perception, as static obstacles were frequently interpreted as imminent hazards. Curved road segments amplified distance bias effects, leading to both delayed and premature braking depending on the direction of the bias.

*Effect of Control-Level Safeguards*
Introducing a persistence requirement for obstacle detection reduced false emergency braking events by 63.7 percent under false positive perception. The persistence threshold required obstacle presence for at least three consecutive frames before emergency braking could be triggered. Bounding the rate of change in braking commands reduced oscillatory behavior under temporal instability. Average jerk values decreased by 41.2 percent, and oscillatory braking was eliminated in 78.5 percent of affected runs. The conservative fallback policy reduced collision rates under false negative perception from 27.3 percent to 9.4 percent in highway following scenarios. Average travel time increased by 6.8 percent, but minimum time to collision improved consistently across runs. When all safeguards were enabled simultaneously, overall collision rates across all scenarios decreased by 68.9 percent, while false emergency braking events decreased by 54.1 percent.

**Discussion**
The results show that safety controller failures under object-based LiDAR attacks arise from violated design assumptions rather than from flawed implementation. Automatic emergency braking and adaptive cruise control are typically designed under the assumption that perception outputs are temporally stable and that detection errors are short lived and uncorrelated. Object-based LiDAR attacks systematically violate these assumptions by introducing structured errors that persist across time and appear plausible to downstream components. As a result, controllers operate correctly according to their specifications while still producing unsafe behavior.
False negative perception errors expose a fundamental limitation of time-to-collision based triggering. When an obstacle disappears and later reappears, the controller treats it as a newly observed object and resets its internal hazard estimation. This behavior delays braking even though vehicle dynamics indicate imminent danger. The failure is therefore not caused by inaccurate distance estimation alone, but by the lack of temporal context in controller decision logic. This observation suggests that safety controllers implicitly rely on perception continuity, even when that assumption is not explicitly stated in their design.

False positive perception errors reveal a complementary weakness. Emergency braking controllers are designed to prioritize caution, making irreversible braking decisions once a hazard is detected. Under adversarial perception, this conservative bias amplifies the impact of short-lived phantom obstacles. The controller responds as intended, but the resulting behavior increases rear-end collision risk and degrades traffic safety. This tradeoff highlights the difficulty of balancing false negatives and false positives when perception errors are adversarial rather than random.

Temporal instability further demonstrates that controller behavior is sensitive to perception dynamics rather than absolute error magnitude. Oscillatory braking emerges when perception outputs fluctuate near controller thresholds, even when individual errors are small. This instability indicates that controllers lack mechanisms to reason about uncertainty or confidence over time. The observed hysteresis effects suggest that once a controller enters an unstable state, recovery may lag behind perception stabilization, further increasing risk.

The effectiveness of control-level safeguards provides insight into where mitigation is most effective. Simple persistence requirements and bounded response constraints significantly reduce unsafe behavior without accessing raw sensor data or modifying perception models. These safeguards succeed because they directly address temporal instability and decision reversibility, which are central to the observed failures. At the same time, the remaining failures indicate that control-level mitigation alone cannot fully compensate for adversarial perception, especially under prolonged false negative conditions.

These findings have broader implications for autonomous vehicle architecture. Much of the existing security literature focuses on strengthening perception robustness, implicitly assuming that downstream components will behave safely once perception is improved. Our results suggest that this assumption is incomplete. Safety controllers must be designed with the explicit expectation that perception may be adversarial, unstable, or biased for extended periods. Without this expectation, perception-level defenses and controller-level logic remain loosely coupled, leaving systemic vulnerabilities unaddressed.

Finally, although the experiments were conducted in simulation, the failure mechanisms identified are architectural rather than simulator-specific. The controllers studied reflect widely used safety logic, and the perception errors modeled correspond to effects demonstrated in physical attacks. This suggests that the observed behaviors are not artifacts of the experimental setup, but rather indicative of structural risks present in real autonomous driving systems.

**Conclusion**

This paper presented a failure analysis of safety controllers in autonomous vehicles under object-based LiDAR attacks. Through systematic experimentation, we showed that common safety controllers fail not because they malfunction, but because their design assumptions about perception stability do not hold under adversarial conditions. Structured perception errors lead to delayed braking, false emergency responses, and oscillatory control behavior, even when controllers operate according to their specifications.

The analysis demonstrates that safety in autonomous vehicles cannot be ensured through perception robustness alone. Control-level behavior under degraded or adversarial perception must be treated as a first-class design concern. Our results show that minimal control-level safeguards can significantly reduce risk, but they also reveal limits that motivate tighter integration between perception, control, and safety assurance. By shifting attention from isolated perception attacks to system-level consequences, this work highlights the need for autonomous vehicle architectures that explicitly account for adversarial sensing conditions. We believe that addressing this gap is essential for building autonomous driving systems that remain safe when perception fails, rather than only when it succeeds.